\documentclass[preprint2]{aastex631}
\usepackage{CJKutf8}
\usepackage{color,soul}
\usepackage{amsmath}

\submitjournal{Proceedings of the National Academy of Sciences on April 22, 2025}

\begin{document}
\begin{CJK*}{UTF8}{gbsn}

\title{Electrical and Thermal Conductivity of Earth's Iron-enriched Basal Magma Ocean}

\correspondingauthor{Francis Dragulet}
\email{francisdragulet@g.ucla.edu}

\author{Francis Dragulet}
\affiliation{Department of Earth, Planetary, and Space Sciences, University of California, Los Angeles, Los Angeles, CA, USA }

\author{Lars Stixrude}
\affiliation{Department of Earth, Planetary, and Space Sciences, University of California, Los Angeles, Los Angeles, CA, USA }



\begin{abstract}
The Earth’s earliest magnetic field may have originated in a basal magma ocean, a layer of silicate melt surrounding the core that could have persisted for billions of years. Recent studies show that the electrical conductivity of liquid with a bulk silicate Earth composition exceeds 10$^{\text{4}}$ S/m at basal magma ocean conditions, potentially surpassing the threshold for dynamo activity. Over most of its history however, the basal magma ocean is more enriched in iron than the bulk silicate Earth, due to iron’s incompatibility in the mineral assemblages of the lower mantle. Using ab-initio molecular dynamics calculations, we examine how iron content affects the silicate dynamo hypothesis. We investigate how the electrical conductivity of silicate liquid changes with iron enrichment, at pressures and temperatures relevant for Earth's basal magma ocean. We also compute the electronic contribution to the thermal conductivity, to evaluate convective instability of basal magma oceans. Finally, we apply our results to model the thermal and magnetic evolution of Earth's basal magma ocean over time.
\end{abstract}



\section{Introduction} \label{sec:intro}

The Earth's magnetic field has been active for at least 3.5 billion years, potentially playing a crucial role in making the planet habitable by shielding the surface from stellar irradiation and preventing atmospheric loss \citep{bono2022pint,brain2024exoplanet}. Today, the magnetic field is sustained by a dynamo process, driven by the convection of liquid iron in the outer core. Given the core's high thermal conductivity, this convection is likely thermochemical, powered by latent heat release and the expulsion of light elements during inner core solidification \citep{pozzo2012thermal, de2012electrical, ohta2016experimental}. Consequently, the age of the core-powered dynamo is closely tied to the inner core's formation, which is estimated to be relatively recent ($<$ 1 Gyr)\citep{labrosse2015thermal, nimmo2015energetics}. Alternative methods of driving convection in the core, such as magnesium precipitation or radioactive heating, may be insufficient to drive a dynamo before inner core nucleation \citep{insixiengmay2025mgo}. This suggests that another mechanism powered the early magnetic field.

The Earth's early dynamo may have been operating in an electrically conducting basal magma ocean (BMO) \citep{ziegler2013implications, stixrude2020silicate}. This layer of silicate melt, located between the core and solid mantle, likely persisted for billions of years, due to the insulating effect of the overlying mantle \citep{labrosse2007crystallizing}. As the BMO slowly crystallized from the top down, its chemical composition evolved through element partitioning between the melt and the overlying solid mantle. Over time, the BMO became significantly enriched in iron, as iron is incompatible with the mineral assemblages of the lower mantle \citep{dragulet2024partitioning, nomura2011spin, tateno2014melting, braithwaite2022partitioning}. This enrichment contributed to its gravitational stability and lowered its melting temperature.

Iron enrichment in the BMO likely influenced its ability to generate a dynamo in two key ways: (1) by increasing the magnetic Reynolds numbers, $R_{m}$, through enhanced electrical conductivity, and (2) by modifying the conditions for convective motion through changes in thermal conductivity. The magnetic Reynolds number, a dimensionless quantity describing the ratio of magnetic induction to magnetic diffusion, is defined as \(R_{m} = \mu_{0} v l \sigma \), where $\mu_{0}$ is the magnetic susceptibility, $v$ is the flow velocity, $l$ is the thickness of the layer, and $\sigma$ is the electrical conductivity. Magnetohydrodynamic simulations suggest that a minimum $R_{m}$ of 40 is required for a self-sustaining dynamo \citep{christensen2006scaling}. However, $R_{m}$ and, in particular, the electrical conductivity of the BMO are poorly constrained. Previous calculations of electrical conductivity have focused on simplified systems - such as MgO, MgSiO$_{3}$, SiO$_{2}$, and bulk silicate Earth liquid \citep{holmstrom2018electronic, scipioni2017electrical, soubiran2018electrical} - that do not capture the iron enrichment expected in a crystallizing BMO.

A silicate dynamo also requires the BMO to be convecting. Convection occurs when the total heat flux out of the BMO, $Q_{\text{total}}$,  exceeds the conductive heat flux $Q_{\text{cond}}$ = $4 \pi r^{2} k \nabla T_{\text{ad}}$, where $r$ is the radius, $k$ is the thermal conductivity, and $\nabla T_{\text{ad}}$ is the adiabatic temperature gradient. If thermal conductivity increases excessively with iron enrichment, heat transport becomes predominantly conductive, which can either completely suppress dynamo action or restrict it to compositional convection alone \citep{driscoll2019geodynamo}. However, the thermal conductivity of the basal magma ocean, and its dependence on iron content, is unknown.

To better constrain the BMO's potential to power a dynamo, we investigate how iron enrichment affects its electrical and thermal conductivity. We perform molecular dynamics simulations of silicate liquid with varying degrees of iron enrichment, quantified by the Fe-Mg fraction: \(X_{\text{Fe}} = \text{Fe}/(\text{Fe}+\text{Mg})\). Our simulations are then combined with Kubo-Greenwood linear response theory to calculate the electronic contributions to electrical and thermal conductivity, following techniques from previous work \citep{holmstrom2018electronic, stixrude2020silicate} - see Methods. We integrate our electrical conductivity results into a thermal evolution model to estimate the time evolution of the magnetic Reynolds number. We find that the electrical conductivity increases significantly with iron content, allowing $R_{m}$ to exceed the threshold for dynamo action in Earth's early history. Although thermal conductivity also rises with iron enrichment, it remains low enough to permit convective motion, suggesting that a silicate dynamo in the BMO was a viable mechanism for powering Earth's early magnetic field.

\section{Results} \label{result}

\subsection{Electrical Conductivity} \label{sigma}

\begin{figure}[t]
\centering
\includegraphics[width=\linewidth]{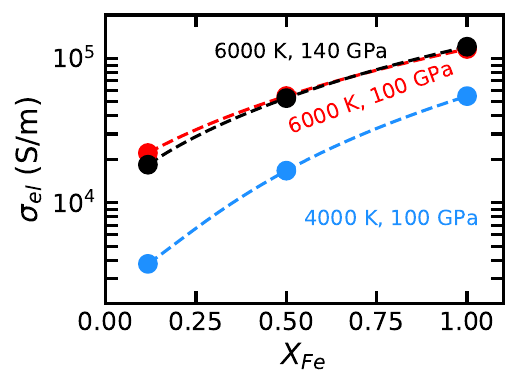}
\caption{Electronic contribution to electrical conductivity $\sigma_{\text{el}}$ of silicate liquid versus iron fraction $X_{\text{Fe}}$ = Fe/(Fe+Mg). Colors indicate different temperature and pressure conditions representative of Earth's basal magma ocean. Circles denote our conductivity results at equilibrium high spin fraction at three iron fractions, while dashed lines represent quadratic fits to these data points. }
\label{fig:sigma_eq}
\end{figure}

Figure \ref{fig:sigma_eq} shows the electronic contribution to the electrical conductivity, $\sigma_{\text{el}}$, for silicate liquid in its equilibrium spin state (see Methods). The silicate liquid composition approximates a pyrolitic mantle \cite{mcdonough1995composition} (Table S1). To assess the impact of iron enrichment, we consider three iron fractions: $X_{\text{Fe}} = \text{Fe}/(\text{Fe}+\text{Mg}) = 0.12$ (pyrolitic), 0.5, and 1, with the latter two more representative of a crystallizing basal magma ocean. We find that $\sigma_{\text{el}}$ is significantly greater than the ionic contribution $\sigma_{\text{ion}}$ (Figure S1). At $X_{\text{Fe}}$ = 0.12, $\sigma_{\text{el}}$ accounts for 70-80 \% of the total electrical conductivity $\sigma_{\text{total}} = \sigma_{\text{el}} + \sigma_{\text{ion}}$, depending on temperature. At $X_{\text{Fe}}$ = 1, $\sigma_{\text{el}}$ contributes more than 90 \% of $\sigma_{\text{total}}$. This indicates that electrons are the dominant charge carriers in iron-bearing silicate liquid for the pressure-temperature-composition conditions examined. 

At the pressures relevant for a basal magma ocean on Earth (100-140 GPa), $\sigma_{\text{el}}$ increases with temperature (Fig.~\ref{fig:sigma_eq}). Along the 6000 K isotherm, $\sigma_{\text{el}}$ decreases slightly with increasing pressure at $X_{\text{Fe}}$ = 0.12. As $X_{\text{Fe}}$ increases, the small effect of pressure is further reduced. Conductivity increases significantly with $X_{\text{Fe}}$: $\sigma_{\text{el}}$ exceeds $10^{5}$ S/m as $X_{\text{Fe}}$ approaches 1, which is roughly an order of magnitude greater than $\sigma_{\text{el}}$ at $X_{\text{Fe}} = 0.12$, yet still below typical metallic conductivities ($>  10^{6}$ S/m).

\begin{figure}
\centering
\includegraphics[width=\linewidth]{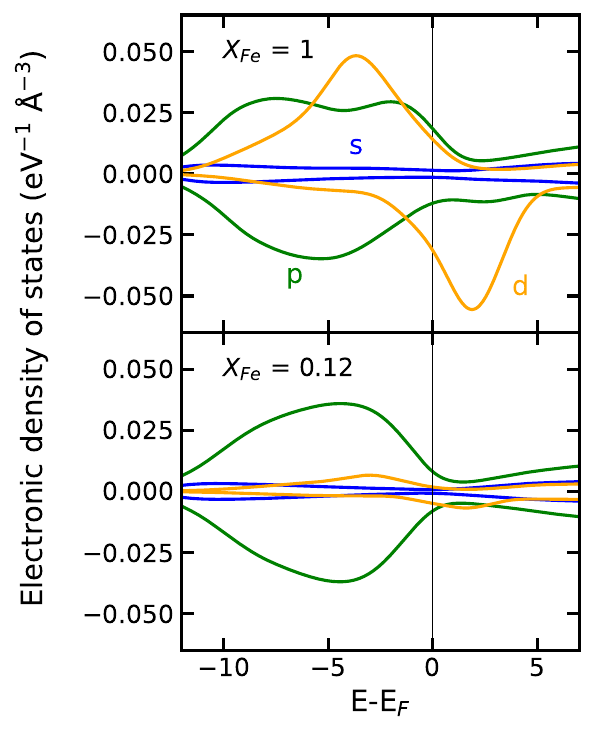}
\caption{Electronic density of states in the high-spin iron bearing silicate liquid at 6000 K and 100$\pm$10 GPa, shown for an iron-rich (top) and iron-poor (bottom) composition. Contributions from s (blue), p (green) and d (orange) states are shown separately, with up-spin and down-spin plotted as positive and negative, respectively. The black vertical line indicates the Fermi energy, $E_{F}$.  }
\label{fig:fermidos}
\end{figure}

We find the origin of the strong dependence of $\sigma_{\text{el}}$ on iron concentration in the electronic density of states. Figure \ref{fig:fermidos} compares the electronic density of states for iron-rich and iron-poor liquids when iron is high-spin, which is the dominant spin state (Figure S2). The conductivity tracks the density of states at the Fermi level, $g(E_{F})$, which is non-zero but lower than typical metals. As $X_{\text{Fe}}$ increases, $g(E_{F})$ rises due to the broad energy bands formed by the 3$d$ electrons of the Fe ions and their hybridization with O 2p states. We find that the increase in $g(E_{F})$ is linear in $X_{\text{Fe}}$ for the pressures, temperatures and magnetic states explored (Figure S3). 

The dependence of $\sigma_{\text{el}}$ on $X_{\text{Fe}}$ is well described by a quadratic function (dotted lines in figure \ref{fig:sigma_eq}). This scaling is consistent with Mott-Ziman theory \cite{mott2012electronic}, which predicts that conductivity is proportional to the square of $g(E_{F})$, which we observe to increase linearly with iron concentration (Figure S3).

\subsection{Thermal Conductivity}

\begin{figure}[t]
\centering
\includegraphics[width=\linewidth]{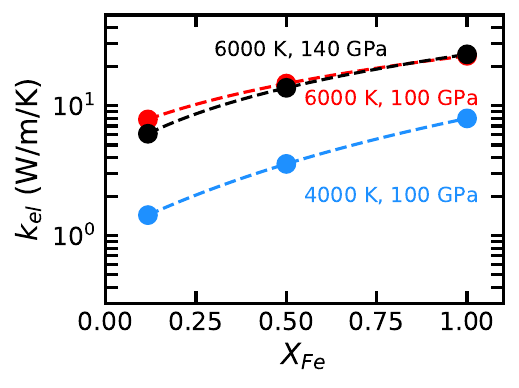}
\caption{Electronic contribution to thermal conductivity $k_{\text{el}}$ of silicate liquid versus iron fraction $X_{\text{Fe}}$ = Fe/(Fe+Mg). As in figure \ref{fig:sigma_eq}, colors indicate temperatures and pressures, while dashed lines represent quadratic fits our conductivity results at the equilibrium spin state (circles).}
\label{fig:kel}
\end{figure}

The electronic contribution to the thermal conductivity, $k_{el}$, in the equilibrium spin state is shown in figure \ref{fig:kel}. While the ionic thermal conductivity, $k_{\text{ion}}$, of iron-enriched silicate liquid is unknown, $k_{\text{ion}}$ of MgSiO$_{3}$ liquid is 4-5 W/m/K at similar pressure and temperature conditions \cite{deng2021thermal} . At low iron concentrations, the electronic thermal conductivity of silicate liquid is comparable to both the ionic contribution in MgSiO$_{\text{3}}$ liquid \cite{deng2021thermal} and the radiative contribution in iron-bearing silicate glasses \cite{murakami2014high}. However, at the higher iron concentrations characteristic of the basal magma ocean, the electronic component is expected to dominate.

The electronic contribution to the thermal conductivity $k_{\text{el}}$ follows similar trends to $\sigma_{\text{el}}$: it is largely insensitive to pressure, increases with temperature, and rises significantly with iron content. As iron concentration increases, $k_{\text{el}}$ remains substantially less than the range calculated for Earth's liquid outer core ($k_{el} > 100$ W/m/K) \cite{pozzo2012thermal, de2012electrical, pozzo2022towards}.

We explore the relationship between the electronic electrical conductivity and thermal conductivity and find that our results do not obey the Wiedemann-Franz law (Figure S4).  This semi-empirical result predicts a linear relationship according to the Lorenz number: $\lambda_{0} = k_{\text{el}} \left(\sigma_{\text{el}} T \right)^{-1} = 2.44 \times 10^{-8}$ $\text{W} \Omega / \text{K}^{2}$. However, this value of the Lorenz number is derived for metals, and we find the Lorenz number in our system to be significantly larger and dependent on pressure, temperature, and composition. Applying the Wiedemann-Franz law to estimate $k_{\text{el}}$ from $\sigma_{\text{el}}$ would lead to an underestimation. As iron concentration increases, the computed Lorenz number approaches the theoretical value, reflecting a trend toward metallic behavior (Figure S4).

Although $k_{\text{el}}$ increases substantially with iron content, it is not high enough to inhibit convection in the magma ocean. Even at the highest calculated values of $k_{\text{el}}$, the conductive heat flux remains less than the total heat flux from the basal magma ocean. For example, taking $k < 30$ W/m/K, an adiabatic temperature gradient of $\nabla T_{\text{ad}}$ of 0.6 K/km \cite{stixrude2009thermodynamics}, and a BMO thickness of 400 km, yields a conductive heat flux, $Q_{\text{cond}} = 4 \pi r^{2} k \nabla T_{\text{ad}}<4$ TW. This is significantly less than the total heat flux out of the basal magma ocean estimated by our thermal evolution model (Figure S5) and by other models \cite{labrosse2007crystallizing, blanc2020thermal}.

\subsection{Thermal and Magnetic Evolution}

\begin{figure}
\centering
\includegraphics[width=\linewidth]{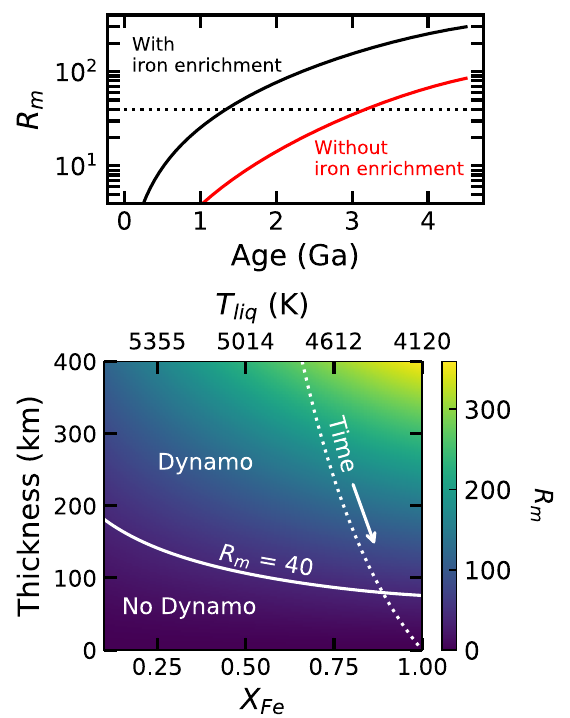}
\caption{\textbf{Top:} Time evolution of the magnetic Reynolds number, $R_{m}$,  as calculated by our thermal evolution model. The red line represents a constant pyrolite composition ($X_{\text{Fe}}$ = 0.12), while the black line accounts for the effect of iron enrichment on electrical conductivity. The dotted line marks the threshold for a self-sustaining dynamo ($R_{\text{m}}>$ 40). \textbf{Bottom:} Parameter space illustrating the effect of basal magma ocean thickness and Fe-Mg fraction $X_{\text{Fe}}$ on the magnetic Reynolds number. The white contour ($R_{m} = 40$) represents the minimum BMO thickness required to sustain a dynamo for a given iron fraction. The dotted white line traces the thickness-$X_{\text{Fe}}$ relationship predicted by our thermal evolution model. The top axis indicates the corresponding liquidus temperature, $T_{\text{liq}}$, defined by $X_{\text{Fe}}$ (equation \ref{evolve2}). $R_{m}$ is calculated using a mixing length velocity scaling.}
\label{fig:rm}
\end{figure}

To assess the potential for a silicate dynamo in Earth's past, we combine our results with a thermal evolution model of the basal magma ocean (Methods). This model tracks the BMO thickness, temperature, and iron content over time, all of which determine the time evolution of conductivity per equations \ref{sigmafit} and \ref{equil_sigma}, along with a quadratic dependence of $\sigma_{\text{el}}$ on $X_{\text{Fe}}$. The model also predicts the heat flux from the BMO, which, together with the conductivity, allows us to calculate the time evolution of the magnetic Reynolds number $R_{m}$. 

Figure \ref{fig:rm} shows the time dependence of the magnetic Reynolds number, assuming a mixing length scaling for flow velocity \cite{christensen2010dynamo} - see Methods. As the basal magma ocean crystallizes and shrinks over time (Figure S6), the magnetic Reynolds number decreases. If the BMO maintains a constant pyrolitic composition (no iron enrichment), $R_{m}$ surpasses the dynamo threshold of ($R_{m}=$ 40) for the first 1.4 billion years of Earth's history. However, accounting for iron enrichment significantly increases the magnetic Reynolds number and the lifetime of the silicate dynamo (black line in figure \ref{fig:rm}). In this case, $R_{m}$ remains above the dynamo threshold for 3.3 billion years, i.e. until 1.2 billion years ago. Using Coriolis-inertial-Archimedean balance scaling for the flow velocity instead reduces the silicate dynamo lifetime by 1.3 billion years (Figure S7).

Also in figure \ref{fig:rm}, we show the dependence of the magnetic Reynolds number on the BMO thickness and $X_{\text{Fe}}$. $X_{\text{Fe}}$ determines the liquidus temperature, $T_{\text{liq}}$, which, along with BMO thickness, controls total heat flowing out of the BMO, and, consequently, the flow velocity (Methods). Increases in $X_{\text{Fe}}$ and BMO thickness both raise $R_{m}$. The minimum BMO thickness required to sustain a dynamo ($R_{m} =$ 40) decreases from 200 to less than 100 km, as $X_{\text{Fe}}$ increases from 0.12 to 1. 

\begin{figure}[h]
\centering
\includegraphics[width=0.85\linewidth]{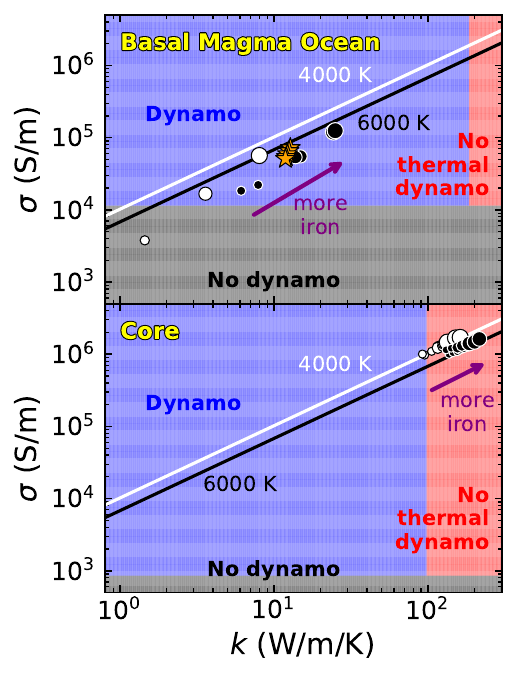}
\caption{Regime diagram illustrating the effect of electrical conductivity $\sigma$ and thermal conductivity $k$ on dynamo production. The boundary between the "dynamo" and "no dynamo" regimes is defined by the magnetic Reynold number \(R_{m} = \mu_{0} v l \sigma  = 40\), while no thermal dynamo will occur if \(Q_{\text{cond}} = 4 \pi r^{2} k \nabla T_{\text{ad} } > Q_{\text{total}}\). The solid lines correspond to the Wiedemann-Franz law relation, \(k_{\text{el}}/\sigma_{\text{el}} = \lambda_{0}T\), at 4000 K (white) and 6000 K (black). \textbf{Top:} basal magma ocean with a thickness of 300 km, total outward heat flux of \(Q_{\text{total}} = 20 \) TW, and adiabatic temperature gradient $\nabla T_{\text{ad}} = 0.6 $ K/km \cite{stixrude2009thermodynamics}.  Circles are electronic conductivity values at 4000 K (white) and 6000 K (black) from figures \ref{fig:sigma_eq} and \ref{fig:kel}, with larger symbols indicating higher iron fraction $X_{\text{Fe}}$. Orange stars represent the time evolution of conductivity predicted by our thermal evolution model. \textbf{Bottom:} liquid core with thickness of 2260 km, \(Q_{\text{total}} = 15 \) TW, and $\nabla T_{\text{ad}} = 1 $ K/km . Circles represent calculations of $\sigma_{\text{el}}$ and $k_{\text{el}}$ for Fe, Fe$_{7}$O, Fe$_{3}$O, Fe$_{7}$Si, and Fe$_{3}$Si liquid from reference \cite{de2012electrical}. These calculations are also along 4000 K (white) and 6000 K (black) isotherms, with larger symbols again corresponding to higher iron content (or lower fraction of light elements). }
\label{fig:regime}
\end{figure}

\section{Discussion}

The basal magma ocean is significantly enriched in iron for most of its lifetime, which has a major impact on its electron transport properties. As the BMO cools, electrical and thermal conductivity tend to decrease with decreasing temperature. However, this tendency is more than compensated by the significant increase in conductivity with increasing iron content, $X_{\text{Fe}}$, as the BMO crystallizes.

The iron-driven increase in conductivity is crucial for sustaining a silicate dynamo. Accounting for iron enrichment raises the magnetic Reynolds number above the threshold required for dynamo action for much longer portion of Earth's history. At the same time, the increase in thermal conductivity with iron enrichment is not sufficient to make conduction the dominant heat transfer mechanism. 

The compositional evolution of the basal magma ocean cannot be solely described by changes in $X_{\text{Fe}}$. In a crystallizing magma ocean, the silica content is expected to diminish with time 
since the liquidus phase at basal magma ocean conditions is bridgmanite \cite{fiquet2010melting, boukare2015thermodynamics, caracas2019melt}. Silica depletion is likely to further enhance $\sigma_{el}$ \cite{holmstrom2018electronic}, potentially allowing the BMO to sustain a dynamo until inner core nucleation, less than 1 billion year ago \cite{labrosse2015thermal,nimmo2015energetics}. 

Furthermore, we have not accounted for compositional convection. Similar to the way in which compositional convection in the core is driven by the accumulation of light elements at the inner core boundary, compositional convection in a BMO can be driven by the descent of iron-enrich liquid during crystallization. This would further enhance the silicate dynamo.

Although the simultaneous increase in electrical and thermal conductivity raises the possibility of reaching a regime where a dynamo driven by thermal convection is not possible\cite{driscoll2019geodynamo}, a silicate dynamo does not reach this limit. This is illustrated in figure \ref{fig:regime}, which shows the effect of $\sigma$ and $k$ on dynamo generation for both the BMO and the core. While the core's thermal conductivity is too high to host a thermal dynamo, the iron-enriched basal magma ocean lies well within the thermal dynamo regime. 

The implications of our findings extend beyond Earth. Other rocky extrasolar planets likely host, or have hosted, basal magma oceans \cite{stixrude2014melting}. This suggests that silicate dynamos may be a widespread phenomenon, potentially playing a critical role in maintaining habitable conditions on rocky planets where a core dynamo is absent. Future work should investigate electrical and thermal conductivities at the more extreme pressures and temperatures of super-Earth basal magma oceans.

\section{Methods} \label{method}

\subsection{Molecular Dynamics Simulations}

Our molecular dynamics simulations are based on density functional theory in the PBEsol approximation \cite{perdew2008restoring} augmented by the "$+U$" method \cite{dudarev1998electron}, with $U-J=2.5$ eV as in our previous work \cite{holmstrom2018electronic, holmstrom2015spin, holmstrom2016spin, stixrude2020silicate}. We utilize the projector augmented plane wave method, as implemented in VASP \cite{kresse1996efficiency, kresse1999ultrasoft}. We perform Born‐Oppenheimer molecular dynamics simulations in the $NVT$ ensemble with periodic boundary conditions, a Nosé‐Hoover thermostat, and a duration of 10–15 ps with 1 fs time step. We assume thermal equilibrium between the ions and electrons via the Mermin functional \cite{mermin1965thermal}. Sampling the Brillouin zone at the Gamma point and a basis‐set energy cutoff of 500 eV converges the energy and pressure to within 3 meV/atom and 0.2 GPa, respectively. We also perform spin-polarized molecular dynamics simulations; for the high-spin simulations, the difference between the number of up-spin and down-spin electrons is set equal to 4 times the number of iron atoms.

Our system contains 149 atoms of six different elements (Mg, Fe, Si, O, Ca and Al), with the relative proportions at $X_{\text{Fe}}=0.12$ chosen to closely match a pyrolite model \cite{mcdonough1995composition} - see table (supplementary table). At higher values of $X_{\text{Fe}}$, Mg is replaced by Fe. 

\subsection{Electrical and Thermal Conductivity}

The electronic contributions to the electrical and thermal conductivity are computed by the Chester-Thellung formulation of the Kubo-Greenwood method:

\begin{equation} \label{sigmael_eq}
    \sigma_{el}(\omega) = L_{1 1}(\omega)
\end{equation}
\begin{equation} \label{kel_eq}
    k_{el}(\omega) = \frac{1}{e^{2}T} \left( L_{22}(\omega) - \frac{L_{12}(\omega) L_{21}(\omega)  }{L_{11}(\omega) } \right)
\end{equation}
The kinetic coefficients $L_{\alpha \beta}(\omega)$ at the electric field frequency $\omega$ are defined as
\begin{equation} \label{onsager}
\begin{split}
    L_{\alpha \beta}(\omega) &= \frac{2 \pi e^{2} \hbar^{2} }{3 m_{e}^{2} \omega \Omega} \sum_{i,j} (f_{i}-f_{j}) \delta(\epsilon_{i} - \epsilon_{j} - \hbar \omega) \\
    & \left| \left< \psi_{i} \left| \nabla \right| \psi_{j} \right> \right|^{2} (-1)^{\alpha + \beta} \left( \epsilon_{i} - \mu \right)^{\alpha - 1} \left( \epsilon_{j} - \mu \right)^{\beta - 1} 
\end{split}
\end{equation}
where the summation is over pairs of states $i,j$, $f$ is the fermi occupation, $\psi$ is the wavefunction, $\epsilon$ is the corresponding single-electron eigenvalue, $\Omega$ is the simulation volume, $m_{e}$ is the electron mass, and $\mu$ is the chemical potential. The values of $\alpha$ and $\beta$ denote whether the electrons are transporting charge or heat. In practice, the $\delta$ function is replaced by a Gaussian with a width given by the average spacing between eigenvalues weighted by the corresponding change in the Fermi function . To obtain the DC conductivity, the frequency dependent conductivity is extrapolated to zero frequency using a linear fit at small $\omega$.

We found both $\sigma_{el}$ and $k_{el}$ to be well converged with a $1 \times 1 \times 1$ k-point mesh and 2500 electronic bands. We compute $\sigma_{el}$, $k_{el}$ and the electronic density of states by averaging over 10 uncorrelated molecular dynamics simulation snapshots.

$\sigma_{el}$ and $k_{el}$ for high-spin and low-spin states are shown in figure S8. For a pyrolitic composition ($X_{\text{Fe}} = 0.12$), the effect of spin polarization is minimal. As $X_{\text{Fe}}$ rises, the difference between high-spin and low-spin conductivities becomes pronounced, with the low-spin state exhibiting larger conductivity. For example, at $X_{\text{Fe}} = 1$ and a temperature of 6000 K, low-spin $\sigma_{\text{el}}$ is at least double that of high-spin $\sigma_{\text{el}}$.

To obtain the total electrical conductivity \(\sigma_{\text{total}} = \sigma_{\text{el}} + \sigma_{\text{ion}}\), we calculate the ionic contribution in the DC limit from the electric current auto correlation function $J(t)$ by
\begin{equation} \label{sigma_ion}
    \sigma_{\text{ion}} = \frac{e^2}{3 k_b T \Omega} \int J(t) dt
\end{equation}
where 
\begin{equation} \label{current_correlation}
    J(t) = \sum_{i,j} z_{i} z_{j} \langle \vec{u}_{i}(t+t_{0}) \cdot \vec{u}_{j}(t_0) \rangle
\end{equation}
The angle brackets indicate an average over time origins $t_0$, and the sum over ions $i$ and $j$ contains the Bader charge $z$ and ion velocity $\vec{u}$. For the total thermal conductivity \(k_{\text{total}}=k_{\text{el}} + k_{\text{ion}}\), we use the ionic thermal conductivity computed for MgSiO$_{\text{3}}$ liquid, 4 W/m/K at BMO conditions \cite{deng2021thermal}. We neglect the radiative contribution to thermal conductivity, $k_{\text{rad}}$, as it is less than 1 W/m/K at the relevant pressures and temperatures \cite{murakami2014high}.

\subsection{Spin Transition}

Equations \ref{sigmael_eq}, \ref{kel_eq} and \ref{onsager} specify the electronic conductivities from the molecular dynamics simulations, which treat iron as either high-spin or low-spin. The conductivity at the equilibrium spin state depends on the fraction of iron atoms that are high-spin, $f_{\text{eq}}$, which we calculate as 
\begin{equation}
    f_{\text{eq}} = \left[ 1 + \exp \left( \frac{\Delta F_{\text{HS-LS}}}{k_{b}T} \right)  \right]^{-1}
\end{equation}
Following our previous work \citep{braithwaite2022partitioning,dragulet2024partitioning}, we compute the free energy difference between liquids with high-spin iron and low-spin iron, $\Delta F_{\text{HS-LS}}$, via thermodynamic integration:
\begin{equation}
    \Delta F_{\text{HS-LS}} = \int^{1}_{0} \langle \Delta U \rangle_{\lambda} d\lambda \approx \frac{\langle \Delta U \rangle_{\lambda = 0} + \langle \Delta U \rangle_{\lambda=1}}{2}
\end{equation}
where $\lambda$ determines the Hamiltonian that produces the molecular dynamics trajectory: $\lambda = 0$ corresponds to the high-spin iron simulation trajectory while $\lambda = 1$ specifies the low-spin trajectory. $\langle \Delta U \rangle$ is the time-averaged difference in the total energy between high-spin and low-spin simulations, calculated by transmuting a low-spin iron atom into a high-spin iron atom. The total energy is $U = E -T(S_{\text{el}} + S_{\text{mag}})$, where $E$ is the internal energy, $T$ is the temperature, $S_{\text{el}}$ is the electronic entropy, and $S_{\text{mag}}$ is the magnetic entropy. The magnetic entropy is $S_{\text{mag}} = Nk_b \text{ln}(\mu_{m}+1)$, with $N$ being the number of iron atoms and $\mu_{m}$ being the magnetic moment averaged over time and number of Fe atoms, in units of Bohr magnetons. Figure S2 shows $f_{\text{eq}}$ and its dependence on pressure, temperature and iron content. 

We find that the high spin is the preferred spin state at lower pressures ($f_{\text{eq}} > 0.8$ at 100 GPa), consistent with other ab-initio studies on liquid silicates and oxides \cite{holmstrom2016spin}, and recent x-ray measurements on shock compressed (Mg$_{0.88}$Fe$_{0.12}$)$_{2}$SiO$_{4}$ liquid \cite{shim2023ultrafast}. As pressure increases, the preferred spin state gradually shifts from high-spin to low-spin. The spin crossover, where $f_{\text{eq}} = 0.5$, depends on $X_{\text{Fe}}$, and occurs within the pressure range of 200-350 GPa for the silicate liquid. 
Notably, the concentration of low-spin iron atoms increases with the iron content, contrasting with the behavior observed in crystalline solids \cite{yoshino2011effect}.

\subsection{Electrical and Thermal Conductivity in Equilibrium}

The high-spin conductivity $\sigma_{\text{el}}^{\text{HS}}$, low-spin conductivity $\sigma_{\text{el}}^{\text{LS}}$, and equilibrium high spin fraction $f_{\text{eq}}$ are used to obtain the equilibrium electrical conductivity $\sigma_{\text{el}}^{\text{eq}}$ via \citep{holmstrom2018electronic}
\begin{equation} \label{equil_sigma}
    \frac{1}{\sigma_{\text{el}}^{\text{eq}}} = \frac{f_{\text{eq}}}{\sigma_{\text{el}}^{\text{HS}}} + \frac{1-f_{\text{eq}}}{\sigma_{\text{el}}^{\text{LS}}}
\end{equation}
and likewise for the equilibrium electronic thermal conductivity:
\begin{equation} \label{equil_k}
    \frac{1}{k_{\text{el}}^{\text{eq}}} = \frac{f_{\text{eq}}}{k_{\text{el}}^{\text{HS}}} + \frac{1-f_{\text{eq}}}{k_{\text{el}}^{\text{LS}}}
\end{equation}

To capture the pressure and temperature dependence of the electrical conductivity, we fit our results to the function
\begin{equation} \label{sigmafit}
    \sigma_{\text{el}} (P,T) = \sigma_{s} \exp \left( - \frac{\Delta E_{s} + P \Delta V_{s}}{R T} \right)
\end{equation}
where $P$ is the pressure, $T$ is the temperature, and $R$ is the gas constant. Similarily, we fit thermal conductivity data points to
\begin{equation} \label{kelfit}
    k_{\text{el}} (P,T) = \sigma_{k} \lambda_{0} T \exp \left( - \frac{\Delta E_{k} + P \Delta V_{k}}{R T} \right)
\end{equation}
where $\lambda_{0} = 2.44 \times 10^{-8}\text{W} \Omega / \text{K}^{2}$ is the theoretical Lorenz number.
We assume the fit parameters $\sigma_{s}$, $\Delta E_{s}$, $\Delta V_{s}$, $\sigma_{k}$, $\Delta E_{k}$, and $\Delta V_{k}$ are independent of pressure and temperature. We perform this fit at each $X_{\text{Fe}}$, with the resulting parameters displayed in Table S2. Furthermore, we fit our conductivity values to all variables to capture the multivariate dependence via a single function: 
\begin{equation} \label{sigmafit_global}
\begin{split}
    \sigma_{\text{el}} (X_{\text{Fe}},P,T) &= \left( \sigma_{u}' + \sigma_{u}''X_{\text{Fe}} + \sigma_{u}'''X_{\text{Fe}}^{2}\right) \\
    & \exp \left( - \frac{\Delta E_{u} + P \Delta V_{u}}{R T} \right)
    \end{split}
\end{equation}
and 
\begin{equation} \label{kel_global}
\begin{split}
    k_{\text{el}} (X_{\text{Fe}},P,T) &= \left( \sigma_{g}' + \sigma_{g}''X_{\text{Fe}} + \sigma_{g}'''X_{\text{Fe}}^{2}\right) \lambda_{0} T \\
    & \exp \left( - \frac{\Delta E_{g} + P \Delta V_{g}}{R T} \right)
\end{split}
\end{equation}
finding $\sigma_{u}'$ = 48298.096 S/m, $\sigma_{u}''$ = 491439.911 S/m, $\sigma_{u}'''$ = 290016.809 S/m, $\Delta E_{u}$ = 101.388 kJ/mol, and $\Delta V_{g}$ = -0.027 cm$^3$/\text{mol}, $\sigma_{g}'$ = 139364.808 S/m, $\sigma_{g}''$ = 559218.167 S/m, $\sigma_{g}'''$ = 122562.441 S/m, $\Delta E_{g}$ = 74.239 kJ/mol, and $\Delta V_{g}$ = 0.035 cm$^3$/\text{mol}.

\subsection{Thermal and Magnetic Evolution}

We model the thermal and magnetic evolution of the basal magma ocean by solving the coupled system of equations:

\begin{equation}\label{evolve1}
\begin{split}
    4 \pi a^{2} k_M \frac{T_{\text{liq}}-T_{M}}{\delta} &= -(M_{m}c_{m}+M_{c}c_{c})\frac{dT_{\text{liq}}}{dt} + H(t)\\
    & - 4 \pi a^{2} \rho \Delta S T_{\text{liq}} \frac{da}{dt}
\end{split}
\end{equation} 

\begin{equation} \label{evolve2}
    T_{\text{liq}} = \left( T_{A}-T_{B} \right)  \left(1- \frac{  ln \left( 1-X_{\text{liq}} \left(1-K_{D} \right) \right)}{ln \left( K_{D} \right)} \right) + T_{B}
\end{equation}

\begin{equation} \label{evolve3}
    \frac{dX_{\text{liq}}}{dt} = - \frac{3a^{2} \left( 1-D_{Fe} \right) X_{\text{liq}}}{a^{3}-b^{3}} \frac{da}{dt}
\end{equation}
Equation \ref{evolve1} describes the heat balance at the top boundary of the BMO, where the left-hand side is the total heat flux out of the BMO. The right-hand side includes contributions from, respectively, BMO secular cooling, core secular cooling, radioactive heat production, and latent heat of freezing. Here, $a$ is the BMO outer radius, $k_M$ is the thermal conductivity of the overlying mantle, $T_\text{liq}$ is the liquidus temperature, $T_M$ is the mantle temperature, $\delta$ is the thickness of the thermal boundary layer above the BMO, $M$ and $c$ are, respectively, the mass and specific heat of the BMO (subscript $m$) and core (subscript $c$), $H$ is the radioactive heat production, $\rho$ is the BMO density, and $\Delta S$ is the entropy change on freezing.

While previous models have assumed a linear phase diagram for $T_{\text{liq}}$ as a function of BMO iron fraction $X_{\text{liq}}$ \cite{labrosse2007crystallizing,stixrude2020silicate, ziegler2013implications}, we adopt a nonlinear phase diagram, whose curvature depends the Fe-Mg distribution coefficient, $K_{\text{D}}$ \cite{stixrude1997structure}. This is equation \ref{evolve2}, where the $T_{\text{liq}}$ also depends on the melting temperature of MgSiO$_3$, $T_A$, and the melting temperature of FeSiO$_3$, $T_B$.

Equation \ref{evolve3} governs the time evolution of $X_{\text{liq}}$, assuming fractional crystallization. The iron partition coefficient $D_{\text{Fe}}$ is related to the distribution coefficient $K_{\text{D}}$ by \(D_{\text{Fe}}=K_{\text{D}} \left(1-X_\text{sol} \right)/ \left(1-X_\text{liq} \right)\), where $X_\text{sol}$ is the iron fraction in the solid phase. $b$ is the radius of the core (3480 km).

We solve these three equations numerically using a fourth-order Runge-Kutta method, yielding the time evolution of the BMO thickness, temperature and iron content (Figures S7 and S8), which in turn allows us to track the time evolution of $\sigma_{\text{total}}$ and $k_{\text{total}}$. We adopt most of the parameters from previous studies \cite{labrosse2007crystallizing,stixrude2020silicate} - see table S3. Exceptions include $T_A$ and $\Delta S$, both of which we take from reference \cite{deng2023melting}, and $T_B$, which we estimate as $T_B = \Delta H/\Delta S$, taking the enthalpy of melting $\Delta H$ from our previous simulations of solid and liquid FeSiO$_{3}$ \cite{dragulet2024partitioning}, and assuming that the entropy of melting $\Delta S$ is the same as that of the MgSiO$_3$ system.

The final component needed to compute the time evolution of the magnetic Reynolds number is the flow velocity. We consider two different velocity scalings \cite{christensen2010dynamo}. The first is mixing length theory (MLT):
\begin{equation}
    v_{\text{MLT}} = \left( \frac{l q}{\rho H_{T}} \right)^{1/3}
\end{equation}
and the second is based on a balance of Coriolis, inertial and Archimedean forces (CIA):
\begin{equation}
    v_{\text{CIA}} = \left( \frac{q}{\rho H_{T}} \right)^{2/5} \left( \frac{l}{\Omega} \right)^{1/5}
\end{equation}
where $q$ is the total heat flow out of the basal magma ocean, $l$ is the BMO thickness, $\rho$ is the BMO density, $H_{T}$ is the thermal scale height \cite{stixrude2009thermodynamics}, and $\Omega$ is the rotation rate.

\section{Acknowledgements} \label{acknowledge}

This work supported by the National Science Foundation under grant EAR-2223935, and computational and storage services of the Hoffman2 Shared Cluster provided by the UCLA Institute for Digital Research and Education's Research Technology Group.





\newpage
\bibliographystyle{elsarticle-harv}
\bibliography{ref}

\begin{thebibliography}{}
\expandafter\ifx\csname natexlab\endcsname\relax\def\natexlab#1{#1}\fi
\providecommand{\url}[1]{\href{#1}{#1}}
\providecommand{\dodoi}[1]{doi:~\href{http://doi.org/#1}{\nolinkurl{#1}}}
\providecommand{\doeprint}[1]{\href{http://ascl.net/#1}{\nolinkurl{http://ascl.net/#1}}}
\providecommand{\doarXiv}[1]{\href{https://arxiv.org/abs/#1}{\nolinkurl{https://arxiv.org/abs/#1}}}

\bibitem[{Blanc {et~al.}(2020)Blanc, Stegman, \& Ziegler}]{blanc2020thermal}
Blanc, N.~A., Stegman, D.~R., \& Ziegler, L.~B. 2020, Earth and Planetary Science Letters, 534, 116085

\bibitem[{Bono {et~al.}(2022)Bono, Paterson, van~der Boon, Engbers, Michael~Grappone, Handford, Hawkins, Lloyd, Sprain, Thallner, {et~al.}}]{bono2022pint}
Bono, R.~K., Paterson, G.~A., van~der Boon, A., {et~al.} 2022, Geophysical Journal International, 229, 522

\bibitem[{Boukar{\'e} {et~al.}(2015)Boukar{\'e}, Ricard, \& Fiquet}]{boukare2015thermodynamics}
Boukar{\'e}, C.-E., Ricard, Y., \& Fiquet, G. 2015, Journal of Geophysical Research: Solid Earth, 120, 6085

\bibitem[{Brain {et~al.}(2024)Brain, Kao, \& O’Rourke}]{brain2024exoplanet}
Brain, D.~A., Kao, M.~M., \& O’Rourke, J.~G. 2024, Reviews in Mineralogy and Geochemistry, 90, 375

\bibitem[{Braithwaite \& Stixrude(2022)}]{braithwaite2022partitioning}
Braithwaite, J., \& Stixrude, L. 2022, Geophysical Research Letters, 49, e2022GL099116

\bibitem[{Caracas {et~al.}(2019)Caracas, Hirose, Nomura, \& Ballmer}]{caracas2019melt}
Caracas, R., Hirose, K., Nomura, R., \& Ballmer, M.~D. 2019, Earth and Planetary Science Letters, 516, 202

\bibitem[{Christensen(2010)}]{christensen2010dynamo}
Christensen, U.~R. 2010, Space science reviews, 152, 565

\bibitem[{Christensen \& Aubert(2006)}]{christensen2006scaling}
Christensen, U.~R., \& Aubert, J. 2006, Geophysical Journal International, 166, 97

\bibitem[{De~Koker {et~al.}(2012)De~Koker, Steinle-Neumann, \& Vl{\v{c}}ek}]{de2012electrical}
De~Koker, N., Steinle-Neumann, G., \& Vl{\v{c}}ek, V. 2012, Proceedings of the National Academy of Sciences, 109, 4070

\bibitem[{Deng {et~al.}(2023)Deng, Niu, Hu, Chen, \& Stixrude}]{deng2023melting}
Deng, J., Niu, H., Hu, J., Chen, M., \& Stixrude, L. 2023, Physical Review B, 107, 064103

\bibitem[{Deng \& Stixrude(2021)}]{deng2021thermal}
Deng, J., \& Stixrude, L. 2021, Geophysical Research Letters, 48, e2021GL093806

\bibitem[{Dragulet \& Stixrude(2024)}]{dragulet2024partitioning}
Dragulet, F., \& Stixrude, L. 2024, Geophysical Research Letters, 51, e2023GL107979

\bibitem[{Driscoll \& Du(2019)}]{driscoll2019geodynamo}
Driscoll, P.~E., \& Du, Z. 2019, Geophysical Research Letters, 46, 7982

\bibitem[{Dudarev {et~al.}(1998)Dudarev, Botton, Savrasov, Humphreys, \& Sutton}]{dudarev1998electron}
Dudarev, S.~L., Botton, G.~A., Savrasov, S.~Y., Humphreys, C., \& Sutton, A.~P. 1998, Physical Review B, 57, 1505

\bibitem[{Fiquet {et~al.}(2010)Fiquet, Auzende, Siebert, Corgne, Bureau, Ozawa, \& Garbarino}]{fiquet2010melting}
Fiquet, G., Auzende, A., Siebert, J., {et~al.} 2010, Science, 329, 1516

\bibitem[{Holmstr{\"o}m \& Stixrude(2015)}]{holmstrom2015spin}
Holmstr{\"o}m, E., \& Stixrude, L. 2015, Physical review letters, 114, 117202

\bibitem[{Holmstr{\"o}m \& Stixrude(2016)}]{holmstrom2016spin}
---. 2016, Physical Review B, 93, 195142

\bibitem[{Holmstr{\"o}m {et~al.}(2018)Holmstr{\"o}m, Stixrude, Scipioni, \& Foster}]{holmstrom2018electronic}
Holmstr{\"o}m, E., Stixrude, L., Scipioni, R., \& Foster, A. 2018, Earth and Planetary Science Letters, 490, 11

\bibitem[{Insixiengmay \& Stixrude(2025)}]{insixiengmay2025mgo}
Insixiengmay, L., \& Stixrude, L. 2025, Earth and Planetary Science Letters, 654, 119242

\bibitem[{Kresse \& Furthm{\"u}ller(1996)}]{kresse1996efficiency}
Kresse, G., \& Furthm{\"u}ller, J. 1996, Computational materials science, 6, 15

\bibitem[{Kresse \& Joubert(1999)}]{kresse1999ultrasoft}
Kresse, G., \& Joubert, D. 1999, Physical review b, 59, 1758

\bibitem[{Labrosse(2015)}]{labrosse2015thermal}
Labrosse, S. 2015, Physics of the Earth and Planetary Interiors, 247, 36

\bibitem[{Labrosse {et~al.}(2007)Labrosse, Hernlund, \& Coltice}]{labrosse2007crystallizing}
Labrosse, S., Hernlund, J., \& Coltice, N. 2007, Nature, 450, 866

\bibitem[{McDonough \& Sun(1995)}]{mcdonough1995composition}
McDonough, W.~F., \& Sun, S.-S. 1995, Chemical geology, 120, 223

\bibitem[{Mermin(1965)}]{mermin1965thermal}
Mermin, N.~D. 1965, Physical Review, 137, A1441

\bibitem[{Mott \& Davis(2012)}]{mott2012electronic}
Mott, N.~F., \& Davis, E.~A. 2012, Electronic processes in non-crystalline materials (OUP Oxford)

\bibitem[{Murakami {et~al.}(2014)Murakami, Goncharov, Hirao, Masuda, Mitsui, Thomas, \& Bina}]{murakami2014high}
Murakami, M., Goncharov, A.~F., Hirao, N., {et~al.} 2014, Nature communications, 5, 5428

\bibitem[{Nimmo(2015)}]{nimmo2015energetics}
Nimmo, F. 2015, Treatise on geophysics, 8, 27

\bibitem[{Nomura {et~al.}(2011)Nomura, Ozawa, Tateno, Hirose, Hernlund, Muto, Ishii, \& Hiraoka}]{nomura2011spin}
Nomura, R., Ozawa, H., Tateno, S., {et~al.} 2011, Nature, 473, 199

\bibitem[{Ohta {et~al.}(2016)Ohta, Kuwayama, Hirose, Shimizu, \& Ohishi}]{ohta2016experimental}
Ohta, K., Kuwayama, Y., Hirose, K., Shimizu, K., \& Ohishi, Y. 2016, Nature, 534, 95

\bibitem[{Perdew {et~al.}(2008)Perdew, Ruzsinszky, Csonka, Vydrov, Scuseria, Constantin, Zhou, \& Burke}]{perdew2008restoring}
Perdew, J.~P., Ruzsinszky, A., Csonka, G.~I., {et~al.} 2008, Physical review letters, 100, 136406

\bibitem[{Pozzo {et~al.}(2012)Pozzo, Davies, Gubbins, \& Alfe}]{pozzo2012thermal}
Pozzo, M., Davies, C., Gubbins, D., \& Alfe, D. 2012, Nature, 485, 355

\bibitem[{Pozzo {et~al.}(2022)Pozzo, Davies, \& Alf{\`e}}]{pozzo2022towards}
Pozzo, M., Davies, C.~J., \& Alf{\`e}, D. 2022, Earth and Planetary Science Letters, 584, 117466

\bibitem[{Scipioni {et~al.}(2017)Scipioni, Stixrude, \& Desjarlais}]{scipioni2017electrical}
Scipioni, R., Stixrude, L., \& Desjarlais, M.~P. 2017, Proceedings of the National Academy of Sciences, 114, 9009

\bibitem[{Shim {et~al.}(2023)Shim, Ko, Sokaras, Nagler, Lee, Galtier, Glenzer, Granados, Vinci, Fiquet, {et~al.}}]{shim2023ultrafast}
Shim, S.-H., Ko, B., Sokaras, D., {et~al.} 2023, Science Advances, 9, eadi6153

\bibitem[{Soubiran \& Militzer(2018)}]{soubiran2018electrical}
Soubiran, F., \& Militzer, B. 2018, Nature communications, 9, 3883

\bibitem[{Stixrude(1997)}]{stixrude1997structure}
Stixrude, L. 1997, Journal of Geophysical Research: Solid Earth, 102, 14835

\bibitem[{Stixrude(2014)}]{stixrude2014melting}
---. 2014, Philosophical Transactions of the Royal Society A: Mathematical, Physical and Engineering Sciences, 372, 20130076

\bibitem[{Stixrude {et~al.}(2009)Stixrude, de~Koker, Sun, Mookherjee, \& Karki}]{stixrude2009thermodynamics}
Stixrude, L., de~Koker, N., Sun, N., Mookherjee, M., \& Karki, B.~B. 2009, Earth and Planetary Science Letters, 278, 226

\bibitem[{Stixrude {et~al.}(2020)Stixrude, Scipioni, \& Desjarlais}]{stixrude2020silicate}
Stixrude, L., Scipioni, R., \& Desjarlais, M.~P. 2020, Nature communications, 11, 935

\bibitem[{Tateno {et~al.}(2014)Tateno, Hirose, \& Ohishi}]{tateno2014melting}
Tateno, S., Hirose, K., \& Ohishi, Y. 2014, Journal of Geophysical Research: Solid Earth, 119, 4684

\bibitem[{Yoshino {et~al.}(2011)Yoshino, Ito, Katsura, Yamazaki, Shan, Guo, Nishi, Higo, \& Funakoshi}]{yoshino2011effect}
Yoshino, T., Ito, E., Katsura, T., {et~al.} 2011, Journal of Geophysical Research: Solid Earth, 116

\bibitem[{Ziegler \& Stegman(2013)}]{ziegler2013implications}
Ziegler, L., \& Stegman, D. 2013, Geochemistry, Geophysics, Geosystems, 14, 4735

\end{thebibliography}


\end{CJK*}
\end{document}